\documentclass[aps,twocolumn,reprint,prl,preprintnumbers,superscriptaddress]{revtex4-1}
\usepackage{amsmath,amssymb,amsthm,mathtools,enumitem,multirow,bm,graphicx,float,array,rotfloat,pgf,soul,bm,comment,enumitem,caption,booktabs,adjustbox,csquotes,mathdots,hyperref,cleveref}
\usepackage{braket,mleftright}
\usepackage[para]{threeparttable}
\usepackage[all]{xy}
\usepackage[normalem]{ulem}

\newcommand{\zmax}{z_{\max}}

\begin{document}

\title{Purely GHZ-like entanglement is forbidden in holography}

\author{Vijay Balasubramanian}
\email[\texttt{vijay@physics.upenn.edu}]{}
\affiliation{Department of Physics and Astronomy, University of Pennsylvania, Philadelphia, PA 19104, U.S.A}
\affiliation{Theoretische Natuurkunde, Vrije Universiteit Brussel and International Solvay Institutes, Pleinlaan 2, B-1050 Brussels, Belgium}
\affiliation{Rudolf Peierls Centre for Theoretical Physics, University of Oxford, Oxford OX1 3PU, U.K.}

\author{Monica Jinwoo Kang}
\email[\texttt{monicak@tamu.edu}]{}
\affiliation{Department of Physics and Astronomy, Texas A\&M University, College Station, TX 77843, U.S.A}

\author{Charlie Cummings}
\email[\texttt{charlie5@sas.upenn.edu}]{}
\affiliation{Department of Physics and Astronomy, University of Pennsylvania, Philadelphia, PA 19104, U.S.A}

\author{Chitraang Murdia}
\email[\texttt{murdia@sas.upenn.edu}]{}
\affiliation{Department of Physics and Astronomy, University of Pennsylvania, Philadelphia, PA 19104, U.S.A}

\author{Simon F. Ross}
\email[\texttt{s.f.ross@durham.ac.uk}]{}
\affiliation{Centre for Particle Theory, Department of Mathematical Sciences, Durham University, Durham DH1 3LE, U.K.}

\begin{abstract}
\noindent 
We provide evidence that three-party entanglement signals in holography obey a relation that is not satisfied by generalized Greenberger-Horne-Zeilinger (GHZ) states. Using proposed holographic duals for these entanglement signals, we provide a geometric argument establishing this relation. This is the first known inequality on the structure of pure three-party holographic states, and shows that time-symmetric holographic states can never have purely GHZ-like entanglement. 
We also discuss similar relations for four parties.  
\end{abstract}

\maketitle
\section{Introduction}

Entanglement is a fundamental feature of quantum systems and has played a central role in recent developments in high-energy and condensed matter physics. In the holographic approach to quantum gravity, strong links between entanglement of the microscopic state and the emergent spacetime geometry have been understood, starting from the seminal conjecture of Ryu and Takayanagi relating von Neumann entropy to minimal surface area in spacetime \cite{Ryu:2006bv}. Many of these developments have focused on bipartite entanglement, where the system is divided into two components. There has also been interest in studying multiparty entanglement and its relation to spacetime in holography. This is particularly natural in the context of geometries with multiple asymptotic regions, such as the multiboundary wormholes considered in \cite{Krasnov:2003ye, Skenderis:2009ju, Balasubramanian:2014hda}, but one can also simply consider dividing a single boundary into multiple pieces. 

The possible structures of multiparty entanglement are extremely rich. Already, when we consider three qubits, there are two inequivalent forms of three-party entanglement, represented by the GHZ and W states, indicating that the three-party entanglement of a state cannot be measured by a single quantity. A general entangled state in three $d$-dimensional systems involves $d^3$ parameters; local unitary transformations on each party only contribute $d^2$ parameters, so the number of inequivalent structures grows rapidly with the size of the parties.

Evidence has been accumulating for the importance of multiparty entanglement in holography. It was conjectured in \cite{Susskind:2014yaa} that GHZ-like entanglement could play an important role in holography. In multiboundary wormholes \cite{Balasubramanian:2014hda}, there are regimes where the dual state is purely bipartite entangled and others where multipartite entanglement is important. The entanglement structures found in these wormhole studies were consistent with a random state model. In \cite{Cui:2018dyq}, it was conjectured that three-party holographic states have mostly bipartite entanglement structure; this was motivated by the observation that a bipartite model sufficed to account for the entanglement entropies of the holographic states across all bipartitions. In \cite{Akers:2019gcv}, a new probe of three-party entanglement between parties $A, B, C$ was devised by taking the difference of the reflected entropy and the mutual information of the reduced density matrix $\rho_{AB}$ on two of the parties, $R^{(3)}(A:B) = S_R(A:B)-I(A:B)$. This has been referred to as the Markov gap \cite{Hayden:2021gno} or the residual information \cite{Balasubramanian:2024ysu}. This is non-vanishing only for states with genuine three-party entanglement. It is non-zero for some holographic states, establishing that they cannot be described by the bipartite model \cite{Umemoto:2019jlz, Akers:2019gcv}. Another probe of three-party entanglement is the genuine multi-entropy  $GM^{(3)}$, which is constructed from the multi-entropy $S^{(3)}$. The multi-entropy $S^{(3)}$ was introduced in  \cite{Gadde:2022cqi} (see also \cite{Penington:2022dhr}). A decomposition of $S^{(3)}$ which removes bipartite contributions was first proposed in \cite{Gadde:2023zzj}, and further developed in \cite{Harper:2024ker}. In \cite{Iizuka:2025ioc} this was systematized to define genuine multi-entropies $GM^{(n)}$ as signals of $n$-party entanglement constructed from the multi-entropies $S^{(n)}$.

In this paper, we continue the study of entanglement properties of holographic states by showing that, assuming conjectured holographic descriptions of the multi-party entanglement signals, there are relations between the multi-party entanglement signals in holography that are not satisfied by general quantum states. This implies restrictions on the entanglement structure in the holographic states. In particular, we show that the holographic descriptions imply that in time-symmetric three-party holographic states, $\frac{1}{2} R^{(3)}(A:B) \geq GM^{(3)}$, where $A,B$ are any two of the parties. This inequality is not satisfied by general quantum states. For example, for systems with three parties, each of which have $d$-dimensional Hilbert spaces, we can define the generalized GHZ states $\ket{\psi_{\text{gGHZ}}} = \sum_{i=1}^d \lambda_i \ket{iii}$, where $\ket{i}$ are an orthonormal basis on each of the Hilbert spaces, and the $\lambda_i$'s are a set of coefficients, satisfying the normalization condition $\sum_{i=1}^{d} |\lambda_i|^2 = 1$. For these states, $R^{(3)} = 0$, while $GM^{(3)} > 0$ \cite{Gadde:2022cqi}. 

We define GHZ-like to mean states that are generalized GHZ up to local unitaries.  If the three parties $A$, $B$ and $C$ are regions of a CFT, so that the total Hilbert space factorizes as ${\cal H} = {\cal H}_A \otimes {\cal H}_B \otimes {\cal H}_C$, then after unitary transformations restricted to the three regions, a GHZ-like state can be written as  $\sum_{i=1}^d \lambda_i \ket{i_A i_B i_C}$. Our inequality establishes that time-symmetric holographic states cannot have only GHZ-like entanglement in this sense (disproving the conjecture of \cite{Susskind:2014yaa}). More generally, if we model the holographic state by a system with many qubits, some of them could have GHZ-like entanglement, but we would then need a sufficient number with other tripartite entanglement structures contributing to $R^{(3)}(A:B)$, so that the bound is satisfied. 

Restrictions on the entanglement structure of holographic states were previously obtained from the holographic entropy cone \cite{Bao:2015bfa}, which gives relations between von Neumann entropies for different bipartitions: for example for mixed states on three parties $A$, $B$, $C$, in holographic states the triple information $I_3 = S(A) +S(B) + S(C) - S(AB) - S(AC) - S(BC) + S(ABC) \leq 0$ \cite{Hayden:2011ag}, while in more general states it can have either sign. (There has been significant further work on holographic entropy inequalities, including \cite{Hubeny:2018ijt, HernandezCuenca:2019wgh, He:2019ttu, Czech:2019lps, Akers:2021lms, Avis:2021xnz, He:2020xuo, Fadel:2021urx, Baiguera:2025dkc, Czech:2021rxe, Hernandez-Cuenca:2023iqh, Czech:2022fzb, Hernandez-Cuenca:2022pst, Czech:2023xed, Czech:2024rco, Bao:2024azn, He:2022bmi, He:2023aif, Grado-White:2024gtx, He:2023rox}. For a review see \cite{Chen:2021lnq}.) This constraint shows that generalized GHZ states with four or more parties do not have semiclassical holographic duals. (An extension of the holographic dictionary to include non-manifold spacetimes was proposed in \cite{Chandra:2022fwi, Jiang:2025iet} to obtain a holographic dual of GHZ states. Our discussion is restricted to bulk spacetimes that are manifolds.)

Our result shows that the holographic entropy cone is only part of the picture of the restrictions on holographic states.  For pure states on three parties, the holographic entropy cone provides no additional restrictions, so it cannot rule out the three-party GHZ state. The three-party GHZ state was ruled out in holography in \cite{Hayden:2021gno}, but their argument did not rely on an explicit inequality. 
Our inequality $\frac{1}{2} R^{(3)}(A:B) \geq GM^{(3)}$ is the first holographic inequality on three-party states. Our derivation holds for time-symmetric spacetimes; we leave the general covariant case for future work. 

In the next section, we review the relevant multiparty entanglement signals and their general properties. We then describe their proposed holographic calculation, and prove the inequality $\frac{1}{2} R^{(3)} \geq GM^{(3)}$. We then discuss higher-party generalizations. 

\section{Tripartite entanglement signals}

We are interested in comparing the residual information $R^{(3)}(A:B) = S_R(A:B)-I(A:B)$, and the genuine multi-entropy $GM^{(3)}$. These are both signals of tripartite entanglement -- functions on a pure three-party state $\ket{\psi}_{ABC}$ that vanish if the state has only bipartite entanglement. Note that the converse is not true; there can be states with genuine tripartite entanglement for which these quantities vanish. For the residual information, this is known to occur for the GHZ state, but there may also be other tripartite entangled states that have $R^{(3)}(A:B)=0$ or $GM^{(3)}=0$. This is why we refer to these as {\em signals} of tripartite entanglement, as opposed to {\em measures}.

For a pure state $\ket{\psi}_{ABC}$, the residual information is obtained by first tracing over one of the parties to obtain a reduced density matrix $\rho_{AB}$. We construct the canonical purification $\ket{\sqrt{\rho}}_{AA^*BB^*}$, and define the reflected entropy \cite{Dutta:2019gen} as the von Neumann entropy of the reduced density matrix on $AA^*$ in this state, 
\begin{equation}
    S_R(A:B) = S_{AA^*}(\ket{\sqrt{\rho}}) \, .  
\end{equation}
This provides a measure of correlation between $A$ and $B$ in the state $\ket{\psi}_{ABC}$. By taking the difference with the mutual information in $\rho_{AB}$, 
\begin{equation}
  I(A:B) = S(A) + S(B) - S(AB) \, ,  
\end{equation}
we can cancel out contributions from purely bipartite entanglement. Thus, the residual information 
\begin{equation}
 R^{(3)}(A:B) = S_R(A:B) - I(A:B) \, ,
\end{equation}
vanishes if $\ket{\psi}_{ABC}$ has only bipartite entanglement \cite{Akers:2019gcv}. Evidently, $R^{(3)}(A:B) \neq 0$ signals tripartite entanglement in the state. Note that this quantity is symmetric, but not permutation invariant; we would obtain independent quantities by tracing out $A$ or $B$ instead and computing $R^{(3)}(B:C)$ or $R^{(3)}(A:C)$. 

The genuine multi-entropy is similarly defined in terms of a quantity called the multi-entropy $S^{(3)}$ \cite{Gadde:2022cqi, Penington:2022dhr}, 
which is sensitive to entanglement between the three parties. The construction of the multi-entropy is somewhat complicated, involving defining a set of R\'enyi multi-entropies $S^{(3)}_n$ and taking the limit as $n \to 1$, so we will not review it here. 
It was shown in \cite{Iizuka:2025ioc} that the difference
\begin{equation}
\begin{aligned}
    GM^{(3)}(A:B:C) =\; & S^{(3)}(A:B:C) \\
    & -\frac{1}{2} \left( S(A) + S(B) + S(C) \right) \, ,
\end{aligned}
\end{equation}
vanishes for states $\ket{\psi}_{ABC}$ with just bipartite entanglement. Thus, this is also a signal of tripartite entanglement. Unlike the residual information, it is permutation invariant on $ABC$, as $S^{(3)}$ is.
Both $R^{(3)}$ and $GM^{(3)}$ are additive under factorization of the state, so they receive no contribution from bipartite entanglement between any two parties.

$R^{(3)}(A:B)$ is non-negative for any quantum state \cite{Dutta:2019gen}. No general statement about the non-negativity of $GM^{(3)}$ is known, but it is non-negative in holographic states \cite{Harper:2024ker, Iizuka:2025ioc}. For the GHZ state \cite{Akers:2019gcv,Balasubramanian:2024ysu,Gadde:2022cqi}, 
\begin{align}\label{eq:GHZ_vals}
    R^{(3)}_{\text{GHZ}} = 0 \,, \quad
    GM^{(3)}_{\text{GHZ}} = \log \sqrt{2}\,.  
\end{align}
Moreover, for the generalized GHZ state \cite{Gadde:2022cqi},
\begin{equation}
    R^{(3)}_{\text{gGHZ}} = 0 \, ,
    \quad
    GM^{(3)}_{\text{gGHZ}}= \sum_{i=1}^{d} |\lambda_i|^2 \log \frac{1}{|\lambda_i|} \,. 
\end{equation}

\section{Tripartite entanglement in holography}

Let us now describe the holographic computation of these quantities. We restrict to a time reflection symmetric spacetime, and compute entanglement at the moment of time symmetry. Evidently, the minimal area surfaces computing these quantities live on the associated time reflection symmetric slice. The full boundary state is taken to be pure, and we divide the boundary into three subregions labeled $A$, $B$, and $C$. 

It is well known that the entanglement entropy is computed by the Ryu-Takayanagi (RT) surface via
\begin{equation}
    S(A) = \frac{\mathcal{A}(\Gamma_A)}{4 G_N} \, ,
\end{equation}
where $\Gamma_A$ is the minimal area surface homologous to $A$ \cite{Ryu:2006bv}.
The bulk region between the boundary subregion $A$ and the RT surface $\Gamma_A$ is called the entanglement wedge $\text{EW}(A)$.

As a consequence of the Ryu-Takayangi formula, the mutual information between $A$ and $B$ in a pure state on three parties $A, B, C$ is given by
\begin{equation}
    I(A:B) = \frac{\mathcal{A}(\Gamma_A) + \mathcal{A}(\Gamma_B) - \mathcal{A}(\Gamma_{C})}{4 G_N} \, ,
\end{equation}
where we have used $S(AB) = S(C)$ as the full system $ABC$ is pure. 

\begin{figure}
\centering
    \includegraphics[scale=0.5]{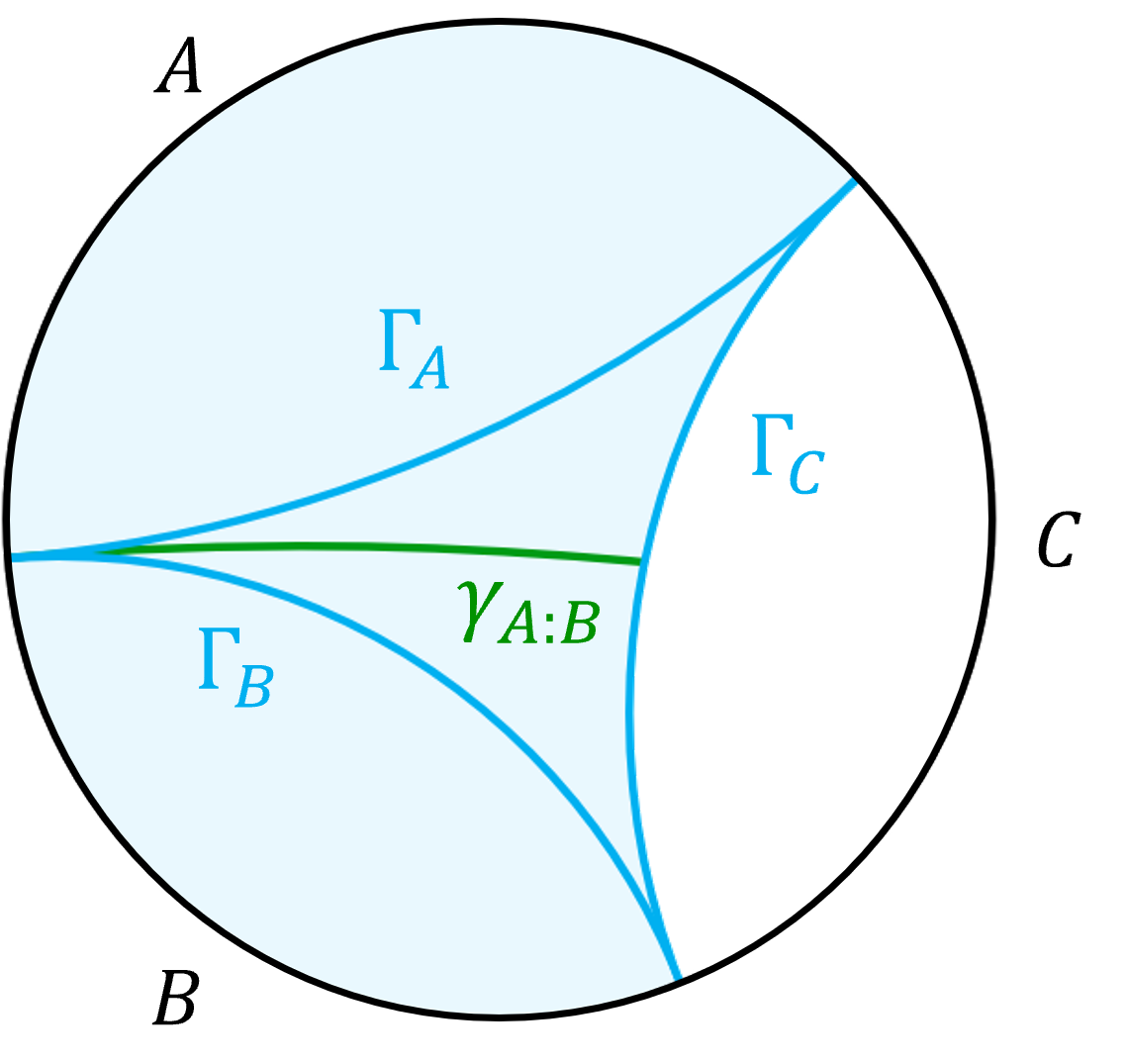}
    \caption{The shaded blue region bounded by the boundary subregion $AB$ and the minimal surface $\Gamma_C = \Gamma_{AB}$ is the entanglement wedge $EW(AB)$. The entanglement wedge cross-section $\gamma_{A:B}$ is the minimal surface that separates $A$ and $B$ within $EW(AB)$.}
    \label{fig:R3_sketch}
\end{figure}

In \cite{Dutta:2019gen}, it was proposed that the reflected entropy is computed by the area of the entanglement wedge cross-section $\gamma_{A:B}$ -- the minimal area surface that separates the subregions $A$ and $B$ within their entanglement wedge $\text{EW}(AB)$,
\begin{equation}
    S_R(A:B) = 2 \frac{\mathcal{A}(\gamma_{A:B})}{4 G_N} \, .
\end{equation}
This is illustrated in an example in vacuum AdS in Figure \ref{fig:R3_sketch}.

\begin{figure}
    \centering
    \includegraphics[scale=0.5]{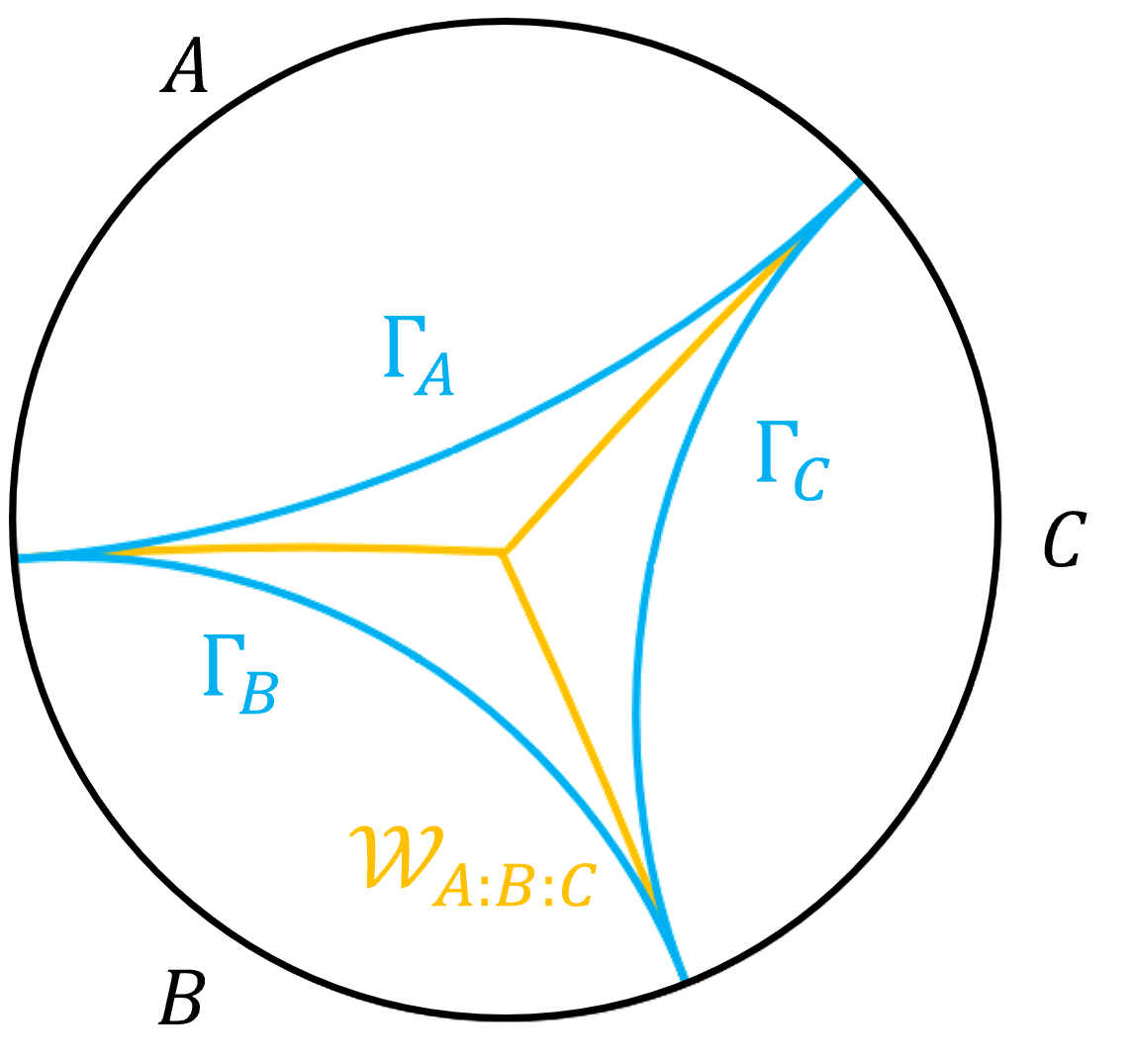}
    \caption{The multi-entropy $S^{(3)}(A:B:C)$ is computed by the minimal brane web $\mathcal{W}_{A:B:C}$ that separates all boundary subregions $A,B,C$ from each other.}
    \label{fig:GM3_sketch}
\end{figure}

In \cite{Gadde:2022cqi}, it was proposed that the multi-entropy $S^{(3)}(A:B:C)$ is computed by the minimal area brane web $\mathcal{W}_{A:B:C}$ which is anchored at the boundaries of the subregions $A$, $B$, and $C$, and  contains sub-webs that are homologous to each of the subregions $A$, $B$, and $C$, 
\begin{equation}
    S^{(3)}(A:B:C) =  \frac{\mathcal{A}(\mathcal{W}_{A:B:C})}{4 G_N} \, . \label{eq:S3_holographic}
\end{equation}
This is illustrated in vacuum AdS in Figure \ref{fig:GM3_sketch}. We will assume this proposal is valid, although there is a subtlety with the analytic continuation to $n = 1$ in calculating the multi-entropy from its R\'enyi versions.\footnote{ In general the continuation from integer $n$ to $n \to 1$ is a subtle question. For the multi-entropy, \cite{Penington:2022dhr} points out that the assumption that the dominant bulk saddle is replica symmetric fails for $n =3$ in the AdS$_3$/CFT$_2$ case. Further discussion of replica symmetry for bulk saddles in the AdS$_3$/CFT$_2$ case is given in \cite{Gadde:2023zzj}. The situation is similar to that in \cite{Belin:2013dva}, where a phase transition in the ordinary Rényi entropy was found as a function of $n$. Consequently, the R\'enyi entropy need not be analytic for all $n$, but only in a neighborhood of $n=1$. Given that a family of replica-symmetric saddles exists for all $n$ and is dominant at sufficiently small $n$ (namely, $n=2$), it is reasonable to similarly conjecture that the replica-symmetric saddles here determine the behavior in the $n \to 1$ limit, although a formal proof would be challenging.} The brane web is a collection of surfaces with possible junctions, satisfying some constraints: in this case, the web must include sub-webs homologous to each of the boundary regions. We minimize the total area of the web subject to these constraints. Depending on the details of the subregions $A,B,C$, a brane web $\mathcal W_{A:B:C}$ may be connected or disconnected, and may or may not contain non-trivial junctions.

Now we proceed to the derivation of the holographic inequality.
Consider the brane web consisting of the minimal surface $\Gamma_C$ and the entanglement wedge cross-section $\gamma_{A:B}$. $\Gamma_C$ separates $AB$ from $C$, while $\gamma_{A:B}$ separates $A$ from $B$ within $AB$. Therefore, the brane web $\Gamma_C \sqcup \gamma_{A:B}$ satisfies the requirements for the brane web $\mathcal{W}_{A:B:C}$. Note that this holds regardless of the topology of each subregion. Therefore, its area must be at least that of the minimal brane web,
\begin{equation}
    \mathcal{A}(\Gamma_C) + \mathcal{A}(\gamma_{A:B}) \geq \mathcal{A}(\mathcal{W}_{A:B:C}) \, .
\end{equation}
Dividing this equation by $4 G_N$ and subtracting $\frac{1}{2}\left(S(A) + S(B) + S(C) \right)$ from both sides, we get the inequality
\begin{equation}
\label{eq:main_1}
    \frac{1}{2} R^{(3)}(A:B) \geq GM^{(3)}(A:B:C) \, .
\end{equation}
This holds in any bulk dimension. Note that the left side of \eqref{eq:main_1} involves a choice of tracing over $C$, while the right side is permutation-invariant. Thus, there are three such inequalities for the three choices of residual information. 

Since $R^{(3)}$ vanishes for GHZ-like states while $GM^{(3)}$ does not, it follows that a holographic state cannot have purely GHZ-like entanglement. Moreover, a holographic state can only contain tensor factors with GHZ-like entanglement if it has sufficient other factors which contribute more strongly to $R^{(3)}$ than to $GM^{(3)}$. 

\begin{figure}
    \centering
    \includegraphics[scale=0.5]{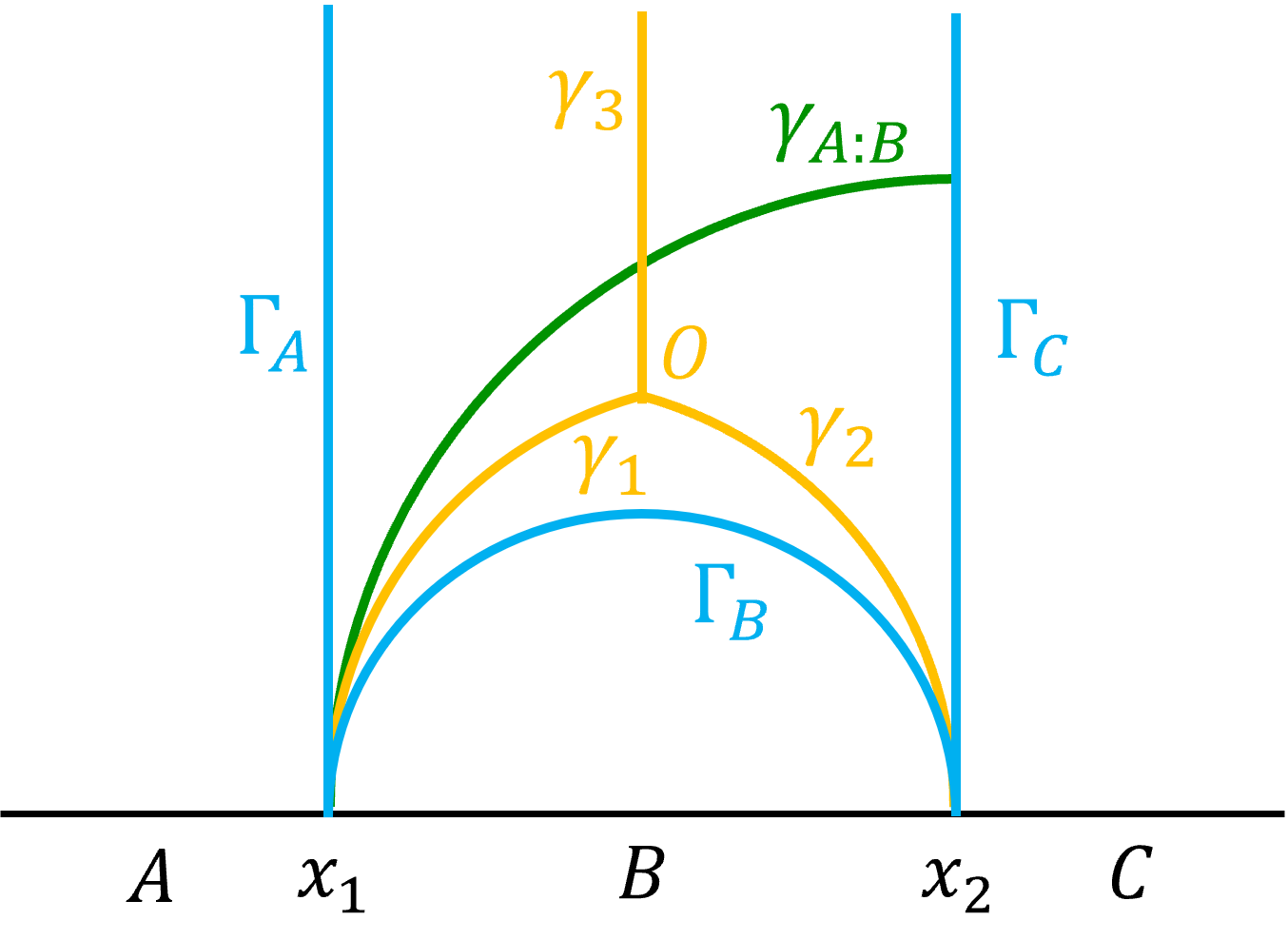}
    \caption{The upper-half plane representation of the hyperbolic disc, showing the minimal surfaces involved in the computation of $R^{(3)}(A:B)$ and the brane web $\mathcal W_{A:B:C}$ involved in the computation of $GM^{(3)}$.}
    \label{fig:R3_GM3_vac}
\end{figure}

We have proved \eqref{eq:main_1} in general, but the argument is abstract, so it may be useful to show that it holds in an explicit example. We therefore consider the calculation of both sides in the vacuum AdS$_3$ case. For $R^{(3)}$ this calculation was performed in \cite{Balasubramanian:2024ysu}; we briefly summarize it here. The constant time slice in vacuum AdS$_3$ is the hyperbolic disc, which we map to the upper-half plane by a conformal transformation. The upper half plane has metric $\mathrm{d}s^2 = \frac{1}{z^2}(\mathrm{d}z^2 + \mathrm{d}x^2)$, where we have set the AdS length scale $\ell_{\text{AdS}} = 1$. We choose a conformal map that sends the boundary point between $A$ and $C$ to infinity. The boundary point between $A$, $B$ and $B$, $C$ lie at $x_1$ and $x_2 > x_1$ respectively, see Fig.~\ref{fig:R3_GM3_vac}. 

The lengths of the surfaces that compute the entanglement entropies are $ \ell_{\Gamma_{A}} = \ell_{\Gamma_{C}} = \log \frac{ \zmax}{\epsilon}$ and $\ell_{\Gamma_{B}} = 2 \log \frac{x_2-x_1}{\epsilon}$, where $\epsilon$ is the radial cutoff and $\zmax$ is the IR cutoff. The reflected entropy is given by twice the length of $\gamma_{A:B}$ which is $\ell_{\gamma} = \log \frac{2(x_2-x_1)}{\epsilon}$.
Therefore, the residual information is \cite{Balasubramanian:2024ysu}
\begin{equation}
\label{eq:R3_vac}
    R^{(3)}(A:B) = \frac{1}{4 G_N} \log 4 \, .
\end{equation}
This is constant (independent of $x_1, x_2$), as expected from conformal invariance, since we could use the conformal transformation to set $x_2 > x_1$ to any values. 

The multi-entropy is given by the total length of the brane web $\mathcal{W}_{A:B:C}$ constructed from the surfaces $\gamma_{1}$, $\gamma_{2}$, and $\gamma_{3}$. We find that $\ell_{\mathcal{W}} = \log \frac{8 (x_2-x_1) \zmax}{3\sqrt{3} \epsilon^2}$, which gives the genuine multi-entropy \cite{Gadde:2023zzj,Harper:2024ker,Iizuka:2025caq}
\begin{equation}
\label{eq:GM3_vac}
    GM^{(3)}(A:B:C) = \frac{3}{4 G_N} \log \frac{2}{\sqrt{3}}  .
\end{equation}
As earlier, this is constant.
Evidently, \eqref{eq:R3_vac} and \eqref{eq:GM3_vac} satisfy the inequality \eqref{eq:main_1}.

\section{Four-party generalization}

In this section, we derive a similar holographic inequality for four parties. There is a four-entropy $S^{(4)}$, defined in \cite{Gadde:2022cqi}. By subtracting a combination of von Neumann entropies from $S^{(4)}$, one can define a one-parameter family of genuine multi-entropies, labelled by a parameter $a$, to obtain a four-party entanglement signal \cite{Iizuka:2025caq}.  The one-parameter ambiguity arises because the triple information $I_3$, which is a combination of von Neumann entropies, is itself a signal of four-party entanglement \cite{Cui:2018dyq}, so this can be included in the four-party genuine multi-entropy with an arbitrary coefficient. Our interest will be in the case labeled  $a=1/3$ in \cite{Iizuka:2025ioc}, 
\begin{equation}
\begin{aligned}
    G&M^{(4)}(A:B:C:D)|_{a=1/3} = S^{(4)}(A:B:C:D) \\
    & - \frac{1}{3} \Big( S^{(3)}(AB:C:D) + S^{(3)}(AC:B:D)  \\ 
    & \qquad + S^{(3)}(AD:B:C) + S^{(3)}(BC:A:D) \\
    &  \qquad + S^{(3)}(BD:A:C) + S^{(3)}(CD:A:B) \Big) \\
    & + \frac{1}{3} \Big( S(AB) + S(AC) + S(BC) \Big).
\end{aligned}
\end{equation}
This particular combination is non-negative in time-symmetric holographic spacetimes \cite{Iizuka:2025caq}.

\begin{figure}
\centering
    \includegraphics[scale=0.5]{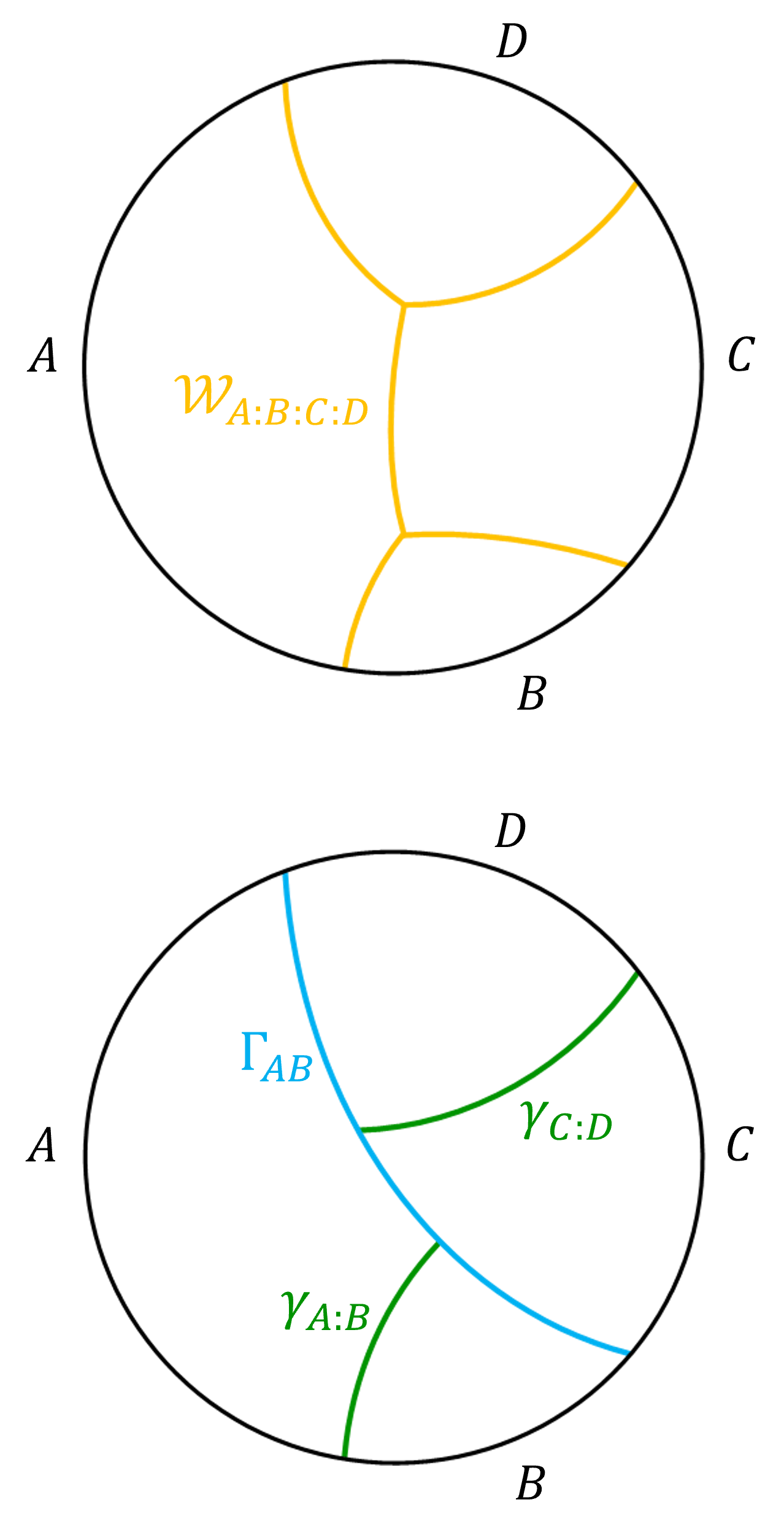}
    \caption{The top figure shows the minimal area brane web $\mathcal{W}_{A:B:C:D}$ that computes $S^{(4)}(A:B:C:D)$. The bottom figure shows a brane web consisting of the  RT surface and the entanglement wedge cross-sections.}
    \label{fig:R3_GM4_sketch}
\end{figure}

In holography, $S^{(4)}(A:B:C:D)$ is again proposed to be given by a minimal area brane web $\mathcal{W}_{A:B:C:D}$ that contains sub-webs homologous to each of the subregions $A$, $B$, $C$ and $D$ \cite{Gadde:2022cqi}. If we consider the union of the RT surface $\Gamma_{AB}$ and the entanglement wedge cross-sections $\gamma_{A:B}$ in the $AB$ entanglement wedge and $\gamma_{C:D}$ in the $CD$ entanglement wedge, this forms a brane web satisfying these constraints. An example of these surfaces is shown in Figure \ref{fig:R3_GM4_sketch}.
Since $\mathcal{W}_{A:B:C:D}$ is the minimal area brane web, we have 
\begin{equation}
    \mathcal{A}(\Gamma_{AB}) + \mathcal{A}(\gamma_{A:B})  + \mathcal{A}(\gamma_{C:D}) \geq \mathcal{A}(\mathcal{W}_{A:B:C:D}) \, .
\end{equation}
It follows that holographic states satisfy the inequality
\begin{equation}
\label{eq:main_2}
\begin{aligned}
    \frac{1}{2} & \Big( R^{(3)}(A:B) \, + \, R^{(3)}(C:D) \Big) \\
    & \geq GM^{(4)}(A:B:C:D) \Big|_{a = 1/3} + \frac{1}{3} \sum GM^{(3)} \\
    & = S^{(4)}(A:B:C:D) - \frac{S(A) + S(B) + S(C) + S(D)}{2} \, ,
\end{aligned}
\end{equation}
where the sum in the middle line runs over all distinct tripartitions of  $ABCD$. It is interesting to note that while the three-party inequality compared three-party entanglement signals, this one compares a combination of three-party signals on the left to a four-party signal on the right. This is somewhat similar to inequalities in the holographic entropy cone, which involve von Neumann entropies for different decompositions of the set of parties. As in the three-party case, the left side is not permutation invariant, so there are actually a set of such inequalities. 

For the four-party GHZ state $\ket{\psi} = \frac{1}{2} (\ket{0000} + \ket{1111})$, the inequality \eqref{eq:main_2} is violated, as $R^{(3)}$ vanishes while $GM^{(3)} = \log \sqrt{2}$ and $GM^{(4)} = 0$ \cite{Iizuka:2025caq}. Since $GM^{(3)}$ is non-negative for holographic states, we also have the simpler inequality for holographic states
\begin{equation}
\label{eq:main_3}
\begin{aligned}
    \frac{1}{2}  \Big( R^{(3)}(A:B) \, & + \, R^{(3)}(C:D) \Big) \\
    & \geq GM^{(4)}(A:B:C:D) \Big|_{a = 1/3} \, .
\end{aligned}
\end{equation}
Note that the four-party GHZ state satisfies this simplified inequality, as both sides vanish.

\begin{figure}[H]
    \centering
    \includegraphics[scale=0.45]{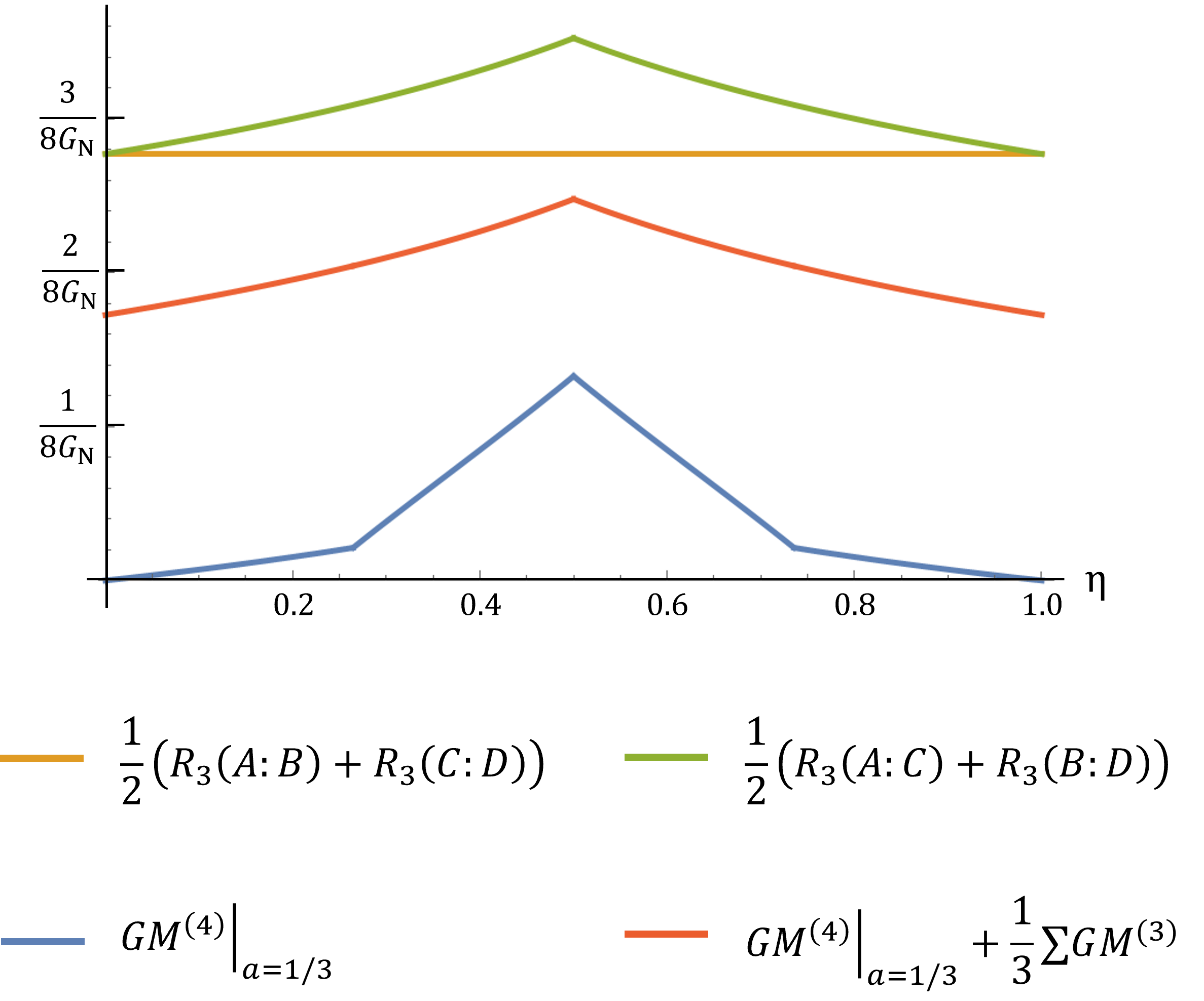}
    \caption{A plot showing the tripartite and quadripartite entanglement quantities for vacuum AdS$_3$ as a function of the conformal cross ratio $\eta$. 
    Clearly, the inequalities \eqref{eq:main_2} and \eqref{eq:main_3} are satisfied.
    }
    \label{fig:GM4_plot}
\end{figure}

As an explicit example, consider vacuum AdS$_3$ with the boundary divided into four connected subregions $A$, $B$, $C$, and $D$.
Due to conformal invariance, any quantity of interest depends only on the conformal cross-ratio $\eta \in (0,1)$ parameterizing the endpoints of the subregions. 
In Figure \ref{fig:GM4_plot}, we plot the two sides of the inequalities in \eqref{eq:main_2} and \eqref{eq:main_3} as a function of this cross-ratio, showing that they are satisfied.

This concludes our analysis for four-party entanglement in holography. An interesting direction for future work is to systematically study other multipartite restrictions for higher numbers of parties.

\section{Acknowledgements}
\noindent M.~J.~K. wishes to thank the organizers and participants of the workshop \emph{Quantum Gravity, Holography and Quantum Information} held in International Institute of Physics, Natal, Brazil, for the stimulating environment and discussions. V.~B., M.~J.~K., and C.~M. were supported in part by the DOE through DESC0013528 and the QuantISED grant DE-SC0020360. V.~B. was supported in part by the Eastman Professorship at Balliol College, University of Oxford. C.~C. is supported by the National Science Foundation Graduate Research Fellowship under Grant No. DGE-2236662.  M.~J.~K.~is supported by the Start-up Research Grant for new faculty provided by Texas A\&M University. S.~F.~R. is supported in part by STFC through grant number ST/X000591/1.

\bibliography{bib}

\end{document}